\documentclass[preprint]{revtex4}%
\usepackage{graphicx}
\usepackage{dcolumn}
\usepackage{bm}
\usepackage{amssymb}
\usepackage{amsmath, amsthm}
\usepackage{color}
\usepackage{setspace}
\usepackage{amsmath}
\usepackage{amsfonts}%
\providecommand{\U}[1]{\protect\rule{.1in}{.1in}}
\begin{document}
\title{Phase Structure of Kerr-AdS Black Hole}
\date{\today}
\author{Tsai, Yu-Dai}
\email{b94201020@ntu.edu.tw}
\affiliation{Department of Physics, National Tsing Hua University, Hsinchu, Taiwan 300, R.O.C.}
\author{X. N. Wu}
\email{wuxn@amss.ac.cn}
\affiliation{Institute of Mathematics, Academy of Mathematics and Systems Science, Chinese
Academy of Sciences, Beijing 100190, P.R.C.}
\author{Yi Yang}
\email{yiyang@mail.nctu.edu.tw}
\affiliation{Department of Electrophysics, National Chiao Tung University, Hsinchu, Taiwan
300, R.O.C.}

\begin{abstract}
We study the critical phenomena of Kerr-AdS black hole. Phase structures are
observed at different temperatures, $T_{L}$, $T_{c1}$ and $T_{c2}$ with
various features. We discuss the thermal stability considering the isothermal
compressibility and how phase transitions related to each other. The
asymptotic value of the angular momentum also has an implication on separating
stable and unstable part. Near critical temperature $T_{c1}$, the order
parameter is determined to calculate the critical exponents. All the critical
exponents ($\alpha$,$\beta$,$\gamma$,$\delta$)=(0,$\frac{1}{2}$,1,3) are
identical to that of mean field systems. We plot the phase diagram near this
critical point, and discuss the scaling symmetry of the free energy.

\end{abstract}
\maketitle

\section{Introduction}

During the last few decades, black hole thermodynamics has been playing the
role of a \textquotedblleft{}thinking experiment\textquotedblright{} to
understand quantum gravity. The discovery of Hawking radiation shows that the
analogy between black hole mechanical laws and the laws of thermodynamics is
physically meaningful. Based on this analogy, Davis pioneered to consider the
phase transition of RN black holes \cite{Davies:1978mf}. Hawking and Page
later investigated the phase transition of Schwarzschild-AdS black holes in
\cite{Hawking:1982dh}. Following their path-breaking research, many works have
been done along this direction and rich phase structures have been discovered
\cite{Lousto:1994jd,
Cai:1996df,Cai:1997cv,Carlip:2003ne,Arcioni:2004ww,Cai:2004pz, Cai:2007vv}.
The later established AdS/CFT duality \cite{Witten:1998qj, Maldacena:1997re}
further inspired people to focus on the asymptotically anti-de Sitter (AdS)
black holes. Critical phenomena were discovered in asymptotically AdS black
holes (see
\cite{Cai:1998ji,Caldarelli:1999xj,Chamblin:1999hg,Wu:2000id,Natsuume:2010vb,Sahay:2010wi,
Banerjee:2010bx,Banerjee:2011gr}).

In \cite{Chamblin:1999hg}, Chamblin et al. studied the phase structures of
RN-AdS black hole. They identified a critical point in RN-AdS black hole by
considering the divergence of heat capacity. Near this critical point, the
behavior of isotherms are similar to that of van der Waals liquid/gas system.
However, the critical exponents of RN-AdS black hole are different from that
of the van der Waals case as it has been shown in \cite{Wu:2000id}. A detailed
investigation of the phase structure of Kerr-AdS black hole is needed to be
compared to the previous results. We hope the report of critical phenomena can
help us to achieve the ultimate goal of finding a microscopic description of
the black hole phase structure.

Another motivation of this work came from the ideas of holographic
superconductors (see \cite{Hartnoll:2008kx,Hartnoll:2008vx}) and their
rotating extension (see \cite{Sonner:2009fk}). In \cite{Sonner:2009fk}, Sonner
studied the superconducting phase transition on the boundary of
Kerr-Newman-AdS black hole. The phase structure of the background field may
affect some properties of the rotating holographic superconductor. And the
knowledge of phase transitions of Kerr-AdS and RN-AdS black holes could be
essential to fully understand the holographic superconductors.

In this paper, we study the phase structure of Kerr-AdS black hole. Rich phase
structure were discovered at three diverse critical temperatures, and this
multi-critical phenomenon in Kerr-AdS black hole has not been carefully
discussed in the previous literatures.

To begin with, we briefly discuss the phase structure by plotting the isotherm
and discribe the three critical temperatures. We then detailedly discuss the
critical behavior of \ each isotherm near the three critical temperature
$T_{L}$, $T_{c1}$ and $T_{c2}$, respectively. We determind the asymptotic
value of the angular momentum, which is important to understand the thermal
stability of the Kerr-AdS black hole.

At a certain temperature $T_{c1}$, we discovered van der Waals-like phase
transition. Unlike the previous case in \cite{Wu:2000id, Sahay:2010wi}, the
critical exponents of Kerr-AdS black hole are found identical to the van der
Waals liquid/gas system and the Weiss ferromagnet. It provides a strong
evidence that Kerr-AdS black hole system belongs to the universality class
which contains these two systems. We also discuss the scaling symmetry of the
free energy near this critical point.

The paper is organized in the following manner. In section \ref{Brief_Intro},
we introduce the necessary basics of Kerr-AdS black hole and define its
thermodynamics quantities. The asymptotic behavior of angular velocity is
briefly discussed. The critical isotherms are plotted in section
\ref{Carving_the_phase_transition}, in which we briefly discuss the phase
structure of Kerr-AdS black hole at $T_{L}$, $T_{c1}$ and $T_{c2}$,
respectively. The asymptotic angular velocity has an implication on the
thermal stability of Kerr-AdS black hole through the phase transition happens
at $T_{c2}$ In section \ref{The_phase_transition_at_Tc1}, we calculate the
critical exponents of the newly discovered critical point at $T_{c1}$ can
compare it to the well-known phase transition systems, such as Weiss
ferromagnet system and van der Waals liquid/gas system. We make analogy of the
free energy and discuss its scaling symmetry. The paper is concluded in
section \ref{Discussion} with discussions and future outlooks.


\section{\label{Brief_Intro}Thermodynamics of Kerr-AdS black hole}

Kerr-AdS black hole is a rotating black hole in AdS space-time. The exact
Kerr-AdS black hole solution of the Einstein equations is given by
\cite{Carter:1968ks} in the Boyer-Lindquist coordinates as%
\begin{equation}
ds^{2}=-\frac{\Delta_{r}}{\Sigma}\left(  dt-\frac{a\sin^{2}\theta}{\Xi}%
d\phi\right)  ^{2}+\frac{\Sigma}{\Delta_{r}}dr^{2}+\frac{\Sigma}%
{\Delta_{\theta}}d\theta^{2}+\frac{\Delta_{\theta}\sin^{2}\theta}{\Sigma
}\left(  adt-\frac{r^{2}+a^{2}}{\Xi}d\phi\right)  ^{2},
\end{equation}
where%
\begin{eqnarray}
\Delta_{r}  &  =\left(  r^{2}+a^{2}\right)  \left(  1+\frac{r^{2}}{l^{2}%
}\right)  -2Mr,\ \ \Xi=1-\frac{a^{2}}{l^{2}},\nonumber\\
\Delta_{\theta}  &  =1-\frac{a^{2}}{l^{2}}\cos^{2}\theta
,\ \ \ \ \ \ \ \ \Sigma=r^{2}+a^{2}\cos^{2}\theta.
\end{eqnarray}
Here $M$ is the mass of Kerr-AdS black hole and $a$ is the rotational
parameter related to the angular momentum of the black hole. If we take $a=0$,
the above metric reduces to the Schwarzschild metric.

The curvature radius $l$ is related to the negative cosmological constant
$\Lambda$ by $\Lambda=-3l^{-2}$. And the radius of horizon $r_{+}$ is defined
by taking $\Delta_{r}=0$, i.e.%
\begin{equation}
(r_{+}^{2}+a^{2})\left(  1+\frac{r_{+}^{2}}{l^{2}}\right)  -2Mr_{+}=0.
\label{r+_a_M}%
\end{equation}
So the area of event horizon $A$ is
\begin{equation}
A=\frac{4\pi(r_{+}^{2}+a^{2})}{\Xi}.
\end{equation}


\subsection{\label{EOS_Main}Thermodynamic Quantities of Kerr-AdS Black Hole}

The thermodynamic quantities of the Kerr-AdS black hole \cite{Gibbons:2004ai}
can be expressed in terms of the radius of horizon $r_{+}$, the rotational
parameter $a$ and the cosmological constant $\Lambda$ ($\Lambda=-3/l^{2}$).
The Hawking temperature of Kerr-AdS black hole is now%
\begin{equation}
T=\frac{3r_{+}^{4}+(a^{2}+l^{2})r_{+}^{2}-l^{2}a^{2}}{4\pi l^{2}r_{+}%
(r_{+}^{2}+a^{2})}, \label{Hawking_T}%
\end{equation}
While the Bekenstein-Hawking entropy is $S=A/4$.

Consider the first law of Kerr-AdS black hole \cite{Gibbons:2004ai}%
\begin{equation}
dE=TdS+\Omega dJ, \label{First_Law}%
\end{equation}
One have to choose the angular velocity $\Omega$ measured relative to a frame
which is non-rotating at infinity. This quantity is defined as%

\begin{equation}
\Omega=\frac{a(1+r_{+}^{2}l^{-2})}{r_{+}^{2}+a^{2}}. \label{Omega_r+_a_Xi}%
\end{equation}
Now for Eq.(\ref{First_Law}), one has the "physical" mass (or energy) $E$ and
angular momentum $J$ defined as%
\begin{equation}
E=\frac{M}{\Xi^{2}},~~J=\frac{Ma}{\Xi^{2}}. \label{mechanic}%
\end{equation}
By solving Eq.(\ref{r+_a_M}) to determine $M$, we have%
\begin{equation}
J=\frac{a(1+\frac{r_{+}^{2}}{l^{2}})(a^{2}+r_{+}^{2})}{2r_{+}(1-\frac{a^{2}%
}{l^{2}})^{2}}. \label{J_r+_a_l}%
\end{equation}

\subsection{\label{EOS_and_TAB}Equation of State and Asymptotic Behavior of
$\Omega$ in Large $J$}

Because $\Lambda$ is a constant, we can rescale each quantity to simplify the
expressions and calculations,%
\begin{eqnarray}
lT & \rightarrow T,&\;l\Omega\rightarrow\Omega,\nonumber\\
\frac{J}{l}  & \rightarrow J&,\;\frac{M}{l}\rightarrow M,\;\frac{a}%
{l}\rightarrow a,\;\frac{r_{+}}{l}\rightarrow r_{+}.\label{rescaling}%
\end{eqnarray}
Thus $J$, $\Omega$, and $T$ become%
\begin{eqnarray}
\label{Equation_of_State}J  
&=&\frac{a(1+r_{+}^{2})(a^{2}+r_{+}^{2})}%
{2r_{+}(1-a^{2})^{2}},\label{J_r+_a}\\
\Omega &=&\frac{a(1+r_{+}^{2})}{r_{+}^{2}+a^{2}},\label{Omega_r+_a}\\
T  & =&\frac{3r_{+}^{4}+a^{2}r_{+}^{2}+r_{+}^{2}-a^{2}}{4\pi r_{+}(r_{+}%
^{2}+a^{2})}.
\end{eqnarray}
Here we have three equations which govern the phase structure of the Kerr-Ads
black hole, which can be regarded as "equations of state". The functions
$\Omega$, $J$ and $T$ are complicatedly dependent on $r_{+}$ and $a$, so that
it is difficult to solve $r_{+}(J,\Omega)$ and $a(J,\Omega)$ analytically to
get the equation of state $T=T(\Omega,J)$ directly.

To begin with, we solve $a$ as a function of $T$ and $r_{+}$ to get%
\begin{equation}
a=\sqrt{\frac{r_{+}^{2}+3r_{+}^{4}-4\pi r_{+}^{3}T}{1-r_{+}^{2}+4\pi r_{+}T}}.
\end{equation}
By putting $a$ back into (\ref{J_r+_a}) and (\ref{Omega_r+_a}), we obtain the
following parametrized expressions for $J$ and $\Omega$,%
\begin{eqnarray}
J  &  =\frac{r_{+}^{2}\sqrt{(1-r_{+}^{2}+4\pi r_{+}T)(1+3r_{+}^{2}-4\pi
r_{+}T)}}{(1-3r_{+}^{2}+4\pi r_{+}T)^{2}},\label{J_in_R,T}\\
\Omega &  =\frac{\sqrt{(1-r_{+}^{2}+4\pi r_{+}T)(1+3r_{+}^{2}-4\pi r_{+}T)}%
}{2r_{+}}. \label{Omega_in_R,T}%
\end{eqnarray}
These two parametrized equations allow us to discuss the asymptotic behavior
of $\Omega$ on an isotherm when $J$ goes to infinity. For a fixed $T$, angular
momentum $J$ goes to infinity as $r_{+}\rightarrow\frac{1}{3}\left(  2\pi
T+\sqrt{3+4\pi^{2}T^{2}}\right)  $. Taking $r_{+}$ to this value, we get%
\begin{equation}
\Omega\rightarrow1.
\end{equation}
Thus we have the asymptotic value of $\Omega$ as one when $J$ goes to infinity.

We name this asymptotic value $\Omega=1$ as $\Omega_{c2}$ because it relates
to the critical temperature $T_{c2}$. This critical behavior at $T_{c2}$ will
be further discussed in section \ref{Critical_Phenomena_of_Tc2}. In the next
section, we will plot the isotherms to study the phase structure of Kerr-AdS
black hole.

\section{\label{Carving_the_phase_transition}Carving the Phase Structure}

\begin{figure}[h]
\includegraphics[width=0.5\textwidth]{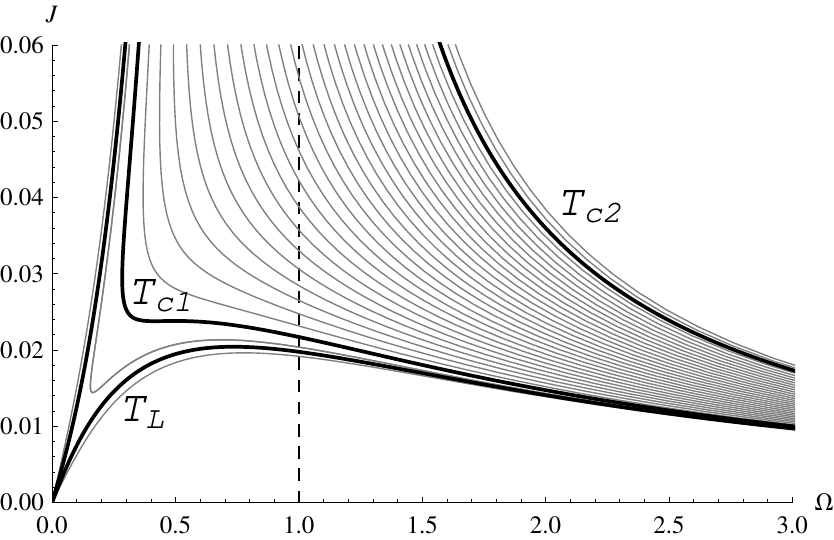}\caption{The
isotherms on the $(J,\Omega)$ space. Three isotherms has been labeled by their
temperatures, $T_{L}$, $T_{c1}$ and $T_{c2}$. $T_{L}>T_{c1}>T_{c2}$.}%
\label{Critical_isotherms}%
\end{figure}

We present the isotherms at various temperatures in FIG.
\ref{Critical_isotherms}. Fos each fixed temperature $T$, we vary $r_{+}$ to
get an isotherm of ($J$, $\Omega$) by using Eq.(\ref{J_in_R,T}) and
Eq.(\ref{Omega_in_R,T}).

The first apparent feature we observed of these isotherms is that above a
certain temperature $T_{L}=0.2757$, the isotherms become $L$-shaped with their
cusp located on $(J,\Omega)=(0,0)$, as one can see in Fig.
\ref{Critical_isotherms}. The Kerr-AdS black hole reduces to Schwarzchild-AdS
black hole at this point. The isotherm has positive slope near the $(0,0)$
point, which makes this part thermally unstable as far as the isothermal
compressibility $\kappa_{T}$ being considered. This $L$-shaping critical
phenomena also destroy the phase structure of $T_{c1}$ at $T_{c1\prime{}}$
($T_{c1}<T_{c1\prime{}}<T_{L}$), as we will discuss in section
\ref{The_phase_transition_at_Tc1}.

There are two other isotherms being specifically labeled in FIG.
\ref{Critical_isotherms} with their temperatures $T_{c1}$ and $T_{c2}$. Two
different critical phenomena occur at lower temperatures. To see the phase
structure at $T_{c1}$ and $T_{c2}$ clearly, we plot the isotherms around the
critical temperatures in different scales in FIG.\ref{Around_Tc1} and
\ref{Log_Diagram}.

\subsection{Phase Structure Near Critical Point at $T_{c1}$}

\begin{figure}[h]
\includegraphics[width=0.5\textwidth]{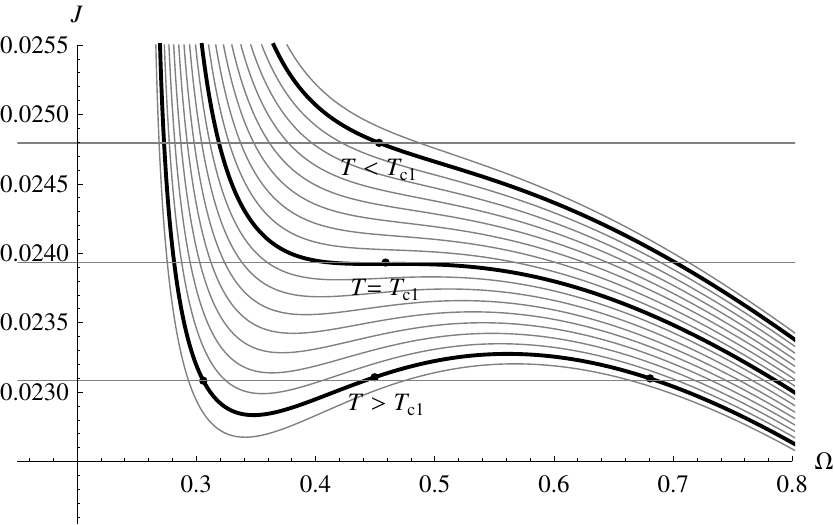}\caption{The
critical isotherms near the critical temperature $T_{c1}$.}%
\label{Around_Tc1}%
\end{figure}

First we discuss the phase transition taking place at $T_{c1}=0.270$. This
temperature is determined by solving the standard critical point equations
(see \cite{Reichl:1980}),%
\begin{equation}
\left(  \frac{\partial J}{\partial\Omega}\right)  _{T_{c1}}=0\text{,
\ \ }\left(  \frac{\partial^{2}J}{\partial\Omega^{2}}\right)  _{T_{c1}}=0.
\end{equation}
By using the above equations of state Eq.(\ref{J_in_R,T}) and
Eq.(\ref{Omega_in_R,T}), we determine $(\Omega_{c1},J_{c1})=(0.459,0.024)$.
The isotherms around this critical point are plotted in FIG. \ref{Around_Tc1}.
At this critical point, the isothermal compressibility $\kappa_{T}=\left(
\frac{\partial{\Omega}}{\partial{J}}\right)  _{T}$ and the specific heat
$C_{J}$ both diverge. And above $T_{c1}$, a single $J$ corresponds to multiple
$\Omega$s for each fixed temperture.

FIG.\ref{Around_Tc1} is similar to the liquid/gas PVT diagram
\cite{Reichl:1980}. This van der Waals-like phase transition can be clearly
visualized as we choose the $J \rightarrow P$, $\Omega\rightarrow V$
correspondence. We will make a detailed study of this critical point in
section \ref{The_phase_transition_at_Tc1}.

\subsection{\label{Critical_Phenomena_of_Tc2}Phase Structure Near Critical
Point at $T_{c2}$}

\begin{figure}[h]
\includegraphics[width=0.8\textwidth]{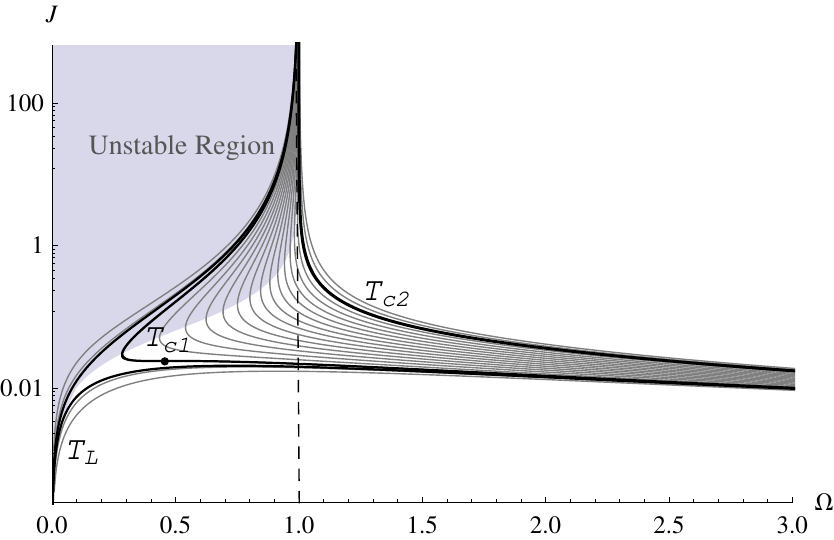}\caption{The diagram
contains all three critical phenomena with their features. Here $J$ is plot on
the log coordinate.}%
\label{Log_Diagram}%
\end{figure}

The other critical temperature, $T_{c2}=0.159$, is determined when an isotherm
has its slope goes to infinity ($\kappa_{T}\rightarrow0$) when $J\rightarrow
\infty$. Below the temperature $T_{c2}$, the isothermal compressibility
$\kappa_{T}$ is always negative along each isotherm, which means that the
black hole is stable under this consideration. But as one raise the
temperature to above $T_{c2}$, positive isothermal compressibility emerge on
the upper part of the curves, making the black hole thermally unstable. We
highlights the unstable region as the shadowed region in FIG.\ref{Log_Diagram}
as well as the critical isotherm at $T_{c2}$. The $J$ value is plotted on the
log coordinate, while $\Omega$ is on the original coordinate, to better
visualize the critical behavior of isotherms near $T_{L}$, $T_{c1}$ and
$T_{c2}$.

One can understand the critical behavior at $T_{c2}$ in another manner. The
asymptotic angular velocity, $\Omega_{c2}$, also marks the critical angular
momentum where the points on the isotherms could become thermally unstable in
the upper part, as we can see in FIG. \ref{Log_Diagram}. On the right side of
$\Omega=\Omega_{c2}$ line, all the isotherms are stable on the upper part. But
on the left side of $\Omega=\Omega_{c2}$, the unstable parts start to emerge
on the upper part of isotherms. It is an important feature: Kerr-AdS black
hole not only has an asymptotic angular velocity, but this asymptotic angular
velocity also determine the thermal stability as far as isothermal
compressibility is considered.

In the end of this section, we summarize the phase structure of Kerr-AdS black
hole with the aid of FIG.\ref{Log_Diagram}. The isotherm at $T_{L}$ has its
cusp located on $(0,0)$. The critical isotherm at $T_{c1}$ has zero slope at
its critical point (the black point in the figure) and then slowly bending
downward as $\Omega$ increases. The isotherm at $T_{c2}$ has infinite slope
when $J$ is very large. One can also see the asymptotic/critical value
$\Omega_{c2}=1$ which separates the stable and the unstable regions in
FIG.\ref{Log_Diagram}.

\section{\label{The_phase_transition_at_Tc1}The Phase Structure near T$_{c1}$}

In this chapter, we consider the phase transition which happens above critical
temperature $T_{c1}$. In FIG. \ref{Around_Tc1}, this van der Waals-like phase
structure can be clearly visualized. This critical point happens when we
choose grand canonical (fixed horizon angular velocity $\Omega$) ensemble. The
critical phenomena has not been discussed carefully in the previous
literatures.As pointed out by \cite{Sahay:2010yq}, the phase structure of
black holes strongly depends on the choice of the ensembles. We study the
quantitative properties of this newly found second-order phase transition, and
compare it to well known systems.


\subsection{Determination of the Critical Point}

The first critical point can be determined by of the following conditions,%
\begin{equation}
\left(  \frac{\partial J}{\partial\Omega}\right)  _{T}=0,\ \ \ \ \ \ \left(
\frac{\partial^{2}J}{\partial\Omega^{2}}\right)  _{T}=0,
\label{critical criteria}%
\end{equation}
which can be expressed as functions of $r_{+}$ and $T$ by Eq.(\ref{J_in_R,T})
and Eq.(\ref{Omega_in_R,T}),%
\begin{eqnarray}
\left(  \frac{\partial J}{\partial\Omega}\right)  _{T}  &  =\frac{\left(
\frac{\partial J}{\partial r_{+}}\right)  _{T}}{\left(  \frac{\partial\Omega
}{\partial r_{+}}\right)  _{T}}=J_{\Omega}(r_{+},T),\\
\left(  \frac{\partial^{2}J}{\partial^{2}\Omega}\right)  _{T}  &
={\frac{\left(  \frac{\left(  \frac{\partial J}{\partial\Omega}\right)  _{T}%
}{\partial r_{+}}\right)  _{T}}{{\left(  \frac{\partial\Omega}{\partial r_{+}%
}\right)  _{T}}}}=J_{\Omega^{2}}(r_{+},T).
\end{eqnarray}
Solving the above two equations, we find that at the critical point
$T=T_{c1}\equiv0.270$ and $r_{+}=r_{c1}\equiv0.459$. Then we obtain
$(\Omega_{c1},J_{c1})=(0.459,0.024)$.

\subsection{Order Parameter and Law of Equal Area}

\begin{figure}[h]
\includegraphics[width=0.5\textwidth]{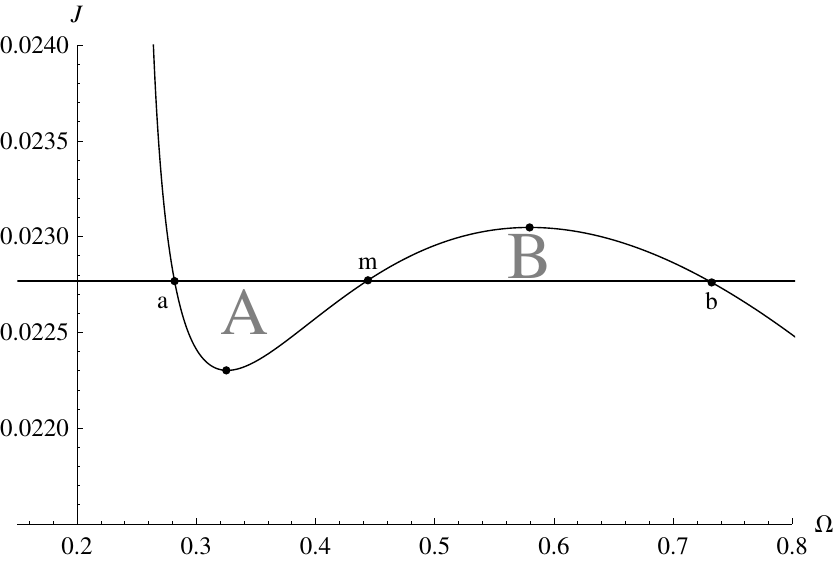}\caption{Applying the
equal-area law on an isotherm at a temperature $T>T_{c1}$.}%
\label{Demo_Equal_Area}%
\end{figure}

We will define the order parameter to describe the critical behavior near the
critical point. Above the critical temperature $T_{c1}$, there are three
points on an isotherm having the same $J$ but with the different $\Omega$s.
Like van der Waals system, one can define an order parameter by Maxwell's
equal-area law. As in FIG. \ref{Demo_Equal_Area}, we choose a isotherm with
$T>T_{c1}$ and draw a horizontal line which intersect the isotherm at three
points $a$,$m$ and $b$. When the area A is equal to the area B, the value
$\eta=(\Omega_{b}-\Omega_{a})/2$ is defined as the order parameter. This
method also allow us to plot the coexistence curve as we present in FIG.
\ref{Coexistence_Curve}. At $T_{c1\prime}=0.2735$ (determined numerically),
the area A of the left part becomes too small to achieve Maxwell's equal-area
law, therefore the phase transition ends at the temperature $T_{c1\prime}$.

For the mean field systems, spontaneous magnetism is the order parameter in
Weiss ferromagnetic system, while the half width of the coexistence curve is
the order parameter in the van der Walls system \cite{Reichl:1980}.

\subsection{\label{Cal_Cri}Critical Exponents}

With the order parameter $\eta=(\Omega_{a}-\Omega_{b})/2$ which we defined in
the previous section, we will calculate the following well known critical
exponents in this section.

\noindent Degree of critical isotherm:%
\begin{equation}
J-J_{c1}=A_{\delta}|\Omega-\Omega_{c1}|^{\delta}sign(\Omega-\Omega
_{c1})\text{, \ \ }T=T_{c1}.
\end{equation}
Degree of coexistence curve:
\begin{equation}
\eta=-A_{\beta}(T-T_{c1})^{\beta},\ \ \ T>T_{c1}.
\end{equation}
Degree of heat capacity ($\Omega=\Omega_{c1})$:
\begin{equation}
C_{\Omega}=\left\{
\begin{array}
[c]{ll}%
A_{\alpha^{\prime}}\{-(T-T_{c1})\}^{-\alpha^{\prime}}, & T<T_{c1}\\
A_{\alpha}\{+(T-T_{c1})\}^{-\alpha}, & T>T_{c1}%
\end{array}
\right.  .
\end{equation}
Degree of isothermal compressibility:
\begin{equation}
\kappa_{T}=\left\{
\begin{array}
[c]{ll}%
A_{\gamma^{\prime}}\{-(T-T_{c2})\}^{-\gamma^{\prime}}, & T<T_{c1}\\
A_{\gamma}\{+(T-T_{c2})\}^{-\gamma}, & T>T_{c1}%
\end{array}
\right.  .
\end{equation}
Following the discussion in \cite{Reichl:1980}, the definitions of the degree
of isothermal compressibility $\kappa_{T}$ for $T<T_{c1}$ and $T<T_{c1}$ are
different. For $T<T_{c1}$, $\gamma^{\prime}$ is defined along the iso-angular
momentum line, i.e., $\Omega=\Omega_{c1}$; when $T>T_{c1}$, $\gamma$ is
defined along the coexistence curve, which is illustrated in FIG.
\ref{Coexistence_Curve}.

Next, we will calculate these critical exponents of the phase transitions in
Kerr-AdS black hole system one by one.

\subsubsection{Degree of Critical Isotherm}

At this critical point $T=T_{c1}$, the first and second derivatives of $J$
over $\Omega$ satisfy%
\begin{equation}
\left.  \left(  \frac{\partial J}{\partial\Omega}\right)  _{T}\right\vert
_{c1}=\left.  \left(  \frac{\partial^{2}J}{\partial\Omega^{2}}\right)
_{T}\right\vert _{c1}=0.
\end{equation}
The third order derivative can be calculated as%
\begin{equation}
\left.  \left(  \frac{\partial^{3}J}{\partial^{3}\Omega}\right)
_{T}\right\vert _{c1}=-0.539\not =0,
\end{equation}
which is not vanishing, therefore we get $\delta=3$.

\subsubsection{\label{Degree of Coexistence Curve}Degree of Coexistence Curve}

\begin{figure}[h]
\includegraphics[width=0.5\textwidth]{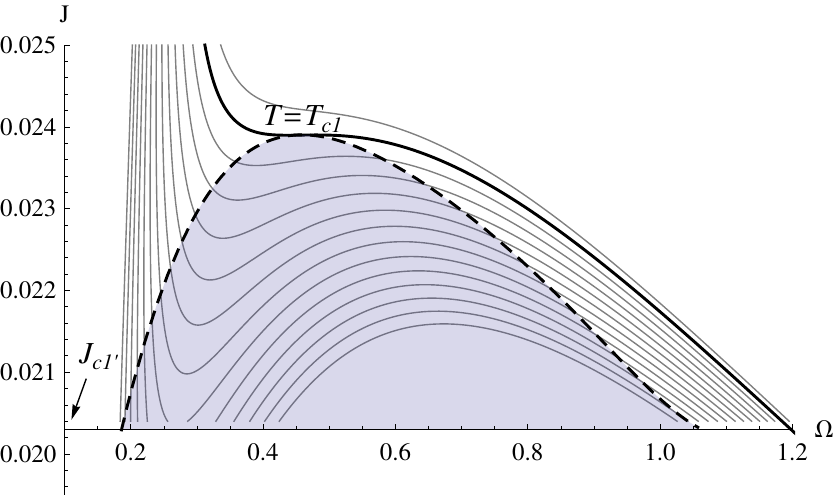}\caption{The
dashed curve is the coexistence curve of phase transition above $T_{c1}$. The
coexistence curve ends at $J_{c1\prime{}}$, because the stable area is too
small to apply the equal area law at the corresponding temperature
$T_{c1\prime{}}$}%
\label{Coexistence_Curve}%
\end{figure}

In FIG. \ref{Coexistence_Curve} we plot the curve of the coexisting states by
Maxwell's equal-area law. This curve indicates the temperature dependence of
the order parameter right after the order parameter emerges.

We expand $J$ in terms of $\Omega$ and $T$ to the third order as%
\begin{eqnarray}
J-J_{c1}  &  \approx\left.  (\partial_{T}J)_{\Omega}\right\vert _{c}%
(T-T_{c1})\nonumber\\
&  +\left.  (\partial_{T}(\partial_{\Omega}J)_{T})_{\Omega}\right\vert
_{c1}(\Omega-\Omega_{c1})(T-T_{c1})\nonumber\\
&  +\frac{1}{2}\left.  (\partial_{T}(\partial_{\Omega}^{2}J)_{T})_{\Omega
}\right\vert _{c1}(\Omega-\Omega_{c})^{2}(T-T_{c1})\nonumber\\
&  +\frac{1}{2}\left.  (\partial_{T}^{2}(\partial_{\Omega}J)_{T})_{\Omega
}\right\vert _{c1}(\Omega-\Omega_{c1})(T-T_{c1})^{2}\nonumber\\
&  +\frac{1}{3!}\left.  (\partial_{\Omega}^{3}J)_{T}\right\vert _{c1}%
(\Omega-\Omega_{c1})^{3}.
\end{eqnarray}
To simplify the calculation, we define%
\begin{equation}
j=J-J_{c1},\ \ \ t=T-T_{c1},\ \ \ \omega=\Omega-\Omega_{c1},
\end{equation}
so that
\begin{equation}
j=c_{10}t+c_{11}\omega t+c_{21}\omega t^{2}+c_{12}\omega^{2}t+c_{03}\omega
^{3}. \label{Expansion_of_J}%
\end{equation}
All coefficients $c_{ij}$ in Eq.(\ref{Expansion_of_J}) can be calculated by
the standard partial differentiation.
The results are
\begin{equation}
c_{10}=-1.027,c_{11}=3.684,c_{12}=-15.792,c_{21}=-458.833,c_{03}=-0.090.
\end{equation}
As we will see in the following discussions, the critical exponents are not
sensitive to the exact values of these constants.

By applying the conditions of equal angular momentum and equal area,%
\begin{equation}
j(\omega_{a},t)=j(\omega_{b},t),\ \ \ \ \ \int_{\omega_{b}}^{\omega_{a}}%
\omega\ dj=0,
\end{equation}
we found that%
\begin{eqnarray}
-(\omega_{a}-\omega_{b})[c_{11}t+c_{21}t^{2}+c_{12}t(\omega_{a}+\omega
_{b})+c_{03}(\omega_{a}^{2}+\omega_{a}\omega_{b}+\omega_{b}^{2})]  &  =0,\\
-(\omega_{a}-\omega_{b})[6c_{11}t(\omega_{a}+\omega_{b})+6c_{21}t^{2}%
(\omega_{a}+\omega_{b})\text{ \ \ \ \ \ \ \ \ \ \ \ \ \ \ \ \ \ \ }  &
\nonumber\\
+8c_{12}t(\omega_{a}^{2}+\omega_{a}\omega_{b}+\omega_{b}^{2})+9c_{03}%
(\omega_{a}^{3}+\omega_{a}^{2}\omega_{b}+\omega_{a}\omega_{b}^{2}+\omega
_{b}^{3})]  &  =0.
\end{eqnarray}
%
Changing the variables as%
\begin{equation}
\omega_{-}\equiv\omega_{b}-\omega_{a}=\Omega_{b}-\Omega_{a},\ \ \omega
_{+}\equiv\omega_{b}+\omega_{a},
\end{equation}
one can solve $\omega_{+}$ and $\omega_{-}$ as%
\begin{eqnarray}
\omega_{+}  &  =-{\frac{2c_{12}t}{3c_{03}}},\\
\omega_{-}  &  =\sqrt{\frac{-4c_{11}t+\left(  \frac{4c_{12}^{2}}{3c_{03}%
}-4c_{21}\right)  t^{2}}{c_{03}}}.
\end{eqnarray}
Now we can expand $t$ in term of $\omega_{-}$ as%
\begin{equation}
t=-\frac{c_{11}c_{03}}{4c_{11}^{2}}\omega_{-}^{2}+O(\omega_{-}^{4}),
\label{beta}%
\end{equation}
to get the order parameter%
\begin{equation}
\frac{\Omega_{b}-\Omega_{a}}{2}\approx A_{\beta}(T-T_{c1})^{\frac{1}{2}}.
\end{equation}
Thus we read $\beta=\frac{1}{2}$.

\subsubsection{Critical Exponent of Heat Capacity}

We now consider the critical exponent of the heat capacity along the constant
angular velocity line $\Omega=\Omega_{c}$. By the black hole thermal dynamical
laws, the role of the internal energy $Q$ is played by the black hole mass $E$
of the Kerr-AdS black hole. $E$ is given by Eq.(\ref{mechanic}) and is
rescaled by Eq(\ref{rescaling}) as%
\begin{equation}
E=\frac{r_{+}^{3}-r_{+}^{5}+4\pi r_{+}^{4}T}{(1-3r_{+}^{2}+4\pi r_{+}T)^{2}}.
\end{equation}
The heat capacity $C_{\Omega}$ can be calculated as
\begin{equation}
C_{\Omega}=\left.  \left(  \frac{\partial E}{\partial T}\right)  _{\Omega
}\right\vert _{c_{1}}=-2.880\not =0.
\end{equation}
Therefore, $\alpha$ and $\alpha^{\prime}$ are both zero because the heat
capacity neither diverges nor vanishes, i.e. $\alpha=\alpha^{\prime}=0$.

\subsubsection{Degree of Isothermal Compressibility}

The isothermal compressibility $\kappa_{T}$ is defined as%
\begin{equation}
\kappa_{T}=\left(  \frac{\partial\Omega}{\partial J}\right)  _{T},
\end{equation}
which is divergent at the critical point. To do the Taylor expansion, we
consider the inverse of the isothermal compressibility%
\begin{equation}
\kappa_{T}^{-1}=\left\{
\begin{array}
[c]{ll}%
A^{\prime}\{-(T-T_{c1})\}^{\gamma^{\prime}}, & (T<T_{c1})\\
A\{+(T-T_{c1})\}^{\gamma}, & (T>T_{c1})
\end{array}
\right.  .
\end{equation}
%
Using Eq. (\ref{Expansion_of_J}), we have%
\begin{equation}
\kappa_{T}^{-1}\propto\left(  \frac{\partial j}{\partial\omega}\right)
_{t}\approx c_{11}t.
\end{equation}
The first order dependence is correct for both $T>T_{c1}$ (along the
iso-angular velocity line, on which $\omega=\omega_{c}$) and $T<T_{c1}$ (along
coexistence curve, on which $\omega\propto t^{\frac{1}{2}}$). Therefore we get
$\gamma=\gamma^{\prime}=1$.


\begin{center}
\begin{table}[ptb]
\caption{The comparison of critical exponents}%
\label{Critical_Table}
\begin{tabular}
[c]{cccccc}\hline
Exponent~~ & Weiss/van der Waals~~ & Kerr-AdS $T_{c1}$~~ & Ising(d=2) & ~
Ising(d=3)~ & \\\hline
$\alpha$ & 0 & 0 & 0 & 0.110(5) & \\
$\beta$ & 1/2 & 1/2 & 1/8 & 0.325+0.0015 & \\
$\gamma$ & 1 & 1 & 7/4 & 1.2405+0.0015 & \\
$\delta$ & 3 & 3 & 15 & 4.82(4) & \\\hline
\end{tabular}
\end{table}
\end{center}


\subsection{\label{Mean_field_comparison}Free Energy and Comparison with Mean
Field Models}

In Table.\ref{Critical_Table}, we compared the critical exponents for various
systems. It is interesting to take the point of view of the mean field theory
to look at the results we got in the previous sections.

In Eq.(\ref{Expansion_of_J}), we expanded the equation of state of Kerr-AdS
black hole in the vicinity of the critical point. With the definition of
$\epsilon=(T-T_{c1})/T_{c1}$ and the order parameter $\eta=(\Omega-\Omega
_{c1})/\Omega_{c1}$, the equation of state becomes
\begin{equation}
j=\tilde{c}_{10}\epsilon+\tilde{c}_{11}\epsilon\eta+\tilde{c}_{21}\epsilon
^{2}\eta+\tilde{c}_{12}\epsilon\eta^{2}+\tilde{c}_{03}\eta^{3},
\label{EOS_Kerr_Eta}%
\end{equation}
where $\tilde{c}_{10}$, $\tilde{c}_{11}$, $\tilde{c}_{21}$, $\tilde{c}_{12}$,
$\tilde{c}_{03}$ are constant coeeficients.

The equation of state can be derived from the free energy,%
\begin{equation}
g(j,\epsilon,\eta)=g_{0}(j,\epsilon)-(j-c_{10}\epsilon)\eta+(c_{11}%
\epsilon+c_{21}\epsilon^{2})\frac{1}{2}\eta^{2}+c_{12}\epsilon\frac{1}{3}%
\eta^{3}+c_{03}\frac{1}{4}\eta^{4}. \label{free_energy_Kerr_AdS}%
\end{equation}
Now we compare our result with some mean field systems.

The first system is Weiss ferromagnet system, see \cite{Goldenfeld:1992qy}.
The equation of state is%
\begin{equation}
\frac{H}{k_{B}T}=M(1-\tau)+M^{3}(\tau-\tau^{2}+\tau^{3}/3+\cdots)~,
\label{Weiss_EOS}%
\end{equation}
where $M$ is the average magnetic moment, $H$ is the external field, and
$\tau^{-1}$ is defined as $T/T_{c}$.

In terms of $\epsilon$ = $(T-T_{c})/T_{c}$ as well as order parameter $\eta
=M$, this equation of state can be written as%
\begin{equation}
\frac{H}{k_{B}T}=\eta\epsilon+\eta^{3}+O(\epsilon\eta^{3})~,
\end{equation}
%
which can be derived from the free energy $\Gamma$ as,%
\begin{equation}
\Gamma(\eta,T,H)=\Gamma_{0}(T,H)-\frac{\eta H}{k_{B}T}+\frac{\epsilon\eta^{2}%
}{2}+\frac{1}{4}\eta^{4}. \label{free_energy_Weiss_system}%
\end{equation}
%
Another mean field system we will consider is van der Waals gas/liquid system,
see also \cite{Goldenfeld:1992qy}. The equation of state is%
\begin{equation}
\pi=\frac{P-P_{c}}{P_{c}}=4\epsilon+6\epsilon\eta+\frac{3}{2}\eta^{3}%
+O(\eta^{4},\eta^{2}\epsilon),
\end{equation}
where $\epsilon=(T-T_{c})/T_{c}$ and the order parameter $\eta=-(V-V_{c}%
)/V_{c}$.

The above equations of state can be derived from the Gibbs free energy $G$,%
\begin{equation}
G(p,T,\eta)=G_{0}(p,T)+\frac{N}{\rho_{0}^{2}}[-(\pi-4\epsilon)\eta
+3\epsilon\eta^{2}+\frac{3}{8}\eta^{4}]. \label{free_energy_Waals}%
\end{equation}
The similarity of Eq.(\ref{free_energy_Kerr_AdS}),
Eq.(\ref{free_energy_Weiss_system}), and Eq.(\ref{free_energy_Waals}) explains
the identical critical exponents in different systems. All systems we
considered above can be incorporated into the classical Landau theory
\cite{Goldenfeld:1992qy}.

\subsection{\label{Widom_Scaling}Widom Scaling}

Now we consider the scaling symmetry of the free energy around the critical
point. We first separate the free energy into two parts,
\begin{equation}
g(\epsilon,j)=g_{r}(\epsilon,j)+g_{s}(\epsilon,j)~.
\end{equation}
Function $g_{r}(\epsilon,j)$ is the regular part which does not change when
approaching the critical point, while $g_{s}(\epsilon,j)$ is the singular part
which possesses the singular behavior of the system in the vicinity of the
critical point.
Assuming that the singular part is a generalized homogeneous function of its
parameters,
through Eq.(\ref{beta}), (\ref{EOS_Kerr_Eta}), and (\ref{free_energy_Kerr_AdS}%
), we can write down the free energy near the critical point as%
\begin{equation}
g_{s}(\epsilon,j)=c_{\epsilon}\epsilon^{2}+c_{j}j^{4/3}~,
\end{equation}
which has the scaling symmetry,%
\begin{equation}
g_{s}(\Lambda^{p}\epsilon,\Lambda^{q}j)=\Lambda g_{s}(\epsilon,j),~~p=\frac
{1}{2},~~q=\frac{3}{4}~.
\end{equation}
Following the discussion of \cite{Reichl:1980}, we can write the critical
exponents in terms of $p$ and $q$ as%
\begin{eqnarray}
\alpha &  =\frac{2p-1}{p}~,\\
\beta &  =\frac{1-q}{p}~,\\
\gamma &  =\frac{2q-1}{p}~,\\
\delta &  =\frac{q}{1-q}. \label{eq:dondon}%
\end{eqnarray}
From the above equations, we find that the critical exponents satisfy the
following expected relations,%
\begin{eqnarray}
&  \alpha+2\beta+\gamma
=2~,\nonumber\label{Scaling_symmetry_of_the_critical_point}\\
&  \alpha+\beta(\delta+1)=2~,\nonumber\\
&  \gamma(\delta+1)=(2-\alpha)(\delta-1)~,\nonumber\\
&  \gamma=\beta(\delta-1).
\end{eqnarray}
The above scaling relations stand as the consistency check for the critical
exponents we obtained in the previous section.

\subsection{\label{Section_Phase_Diagram} Phase Diagram}

\begin{figure}[h]
\includegraphics[width=0.5\textwidth]{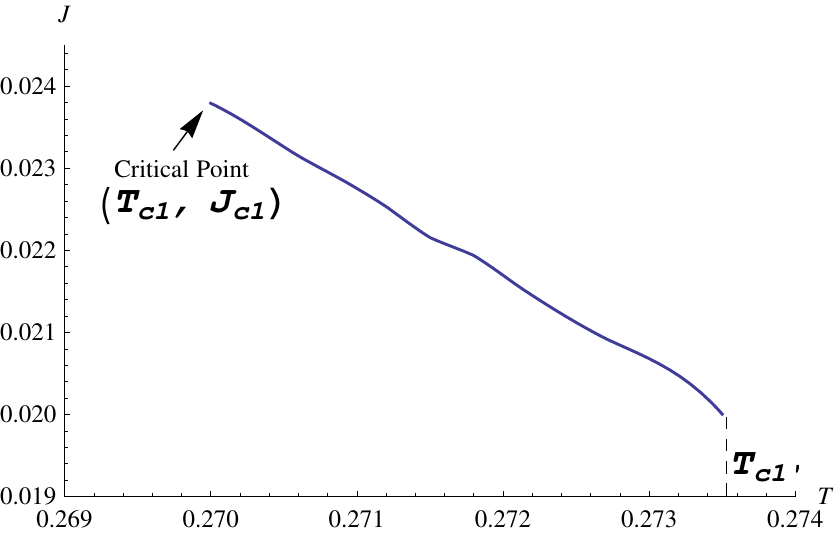}\caption{Phase
diagram near critical point $T_{c1}$. Similar to the phase diagram of
liquid/gas phase transition\cite{Reichl:1980,Stanley:1987}.}%
\label{Phase_Diagram_KerrAdS}%
\end{figure}

In FIG \ref{Phase_Diagram_KerrAdS}, we present the phase diagram near critical
point $(T_{c1},J_{c1})$ on the $T-J$ plane. One can see the phase transition
line and the critical point. Liquid/gas-like phase transition occurs when
crossing the phase transition line. This phase structure occurs at the
temperature above $T_{c1}$, and ends when the temperature reaches
$T_{c1\prime{}}$, at which the Maxwell's equal-area law fails to apply on the
isotherm due to the stable part is too small.

Although this phase diagram is similar to that of liquid/gas phase transition,
this critical phenomenon happens when raising the temperature above $T_{c1}$,
rather than lowering the temperature below the critical temperature. The phase
transition line thus point to different direction as that of liquid/gas system
\cite{Reichl:1980,Stanley:1987}. This is an intriguing feature of van der
Waals-like critical point for both RN-AdS black hole \cite{Wu:2000id,
Sahay:2010wi} and Kerr-AdS black hole.

\section{\label{Discussion}Discussion}

In this paper, we studied the phase structure of Kerr-AdS black hole. Rich
critical phenomena have been found at the different temperatures $T_{L}$,
$T_{c1}$ and $T_{c2}$. Based on the isotherms, we discussed the physical
meaning of each critical point. The asymptotic angular momentum $\Omega_{c2}$
was also discussed and found to be assoiated to the critical behavior at
$T_{c2}$. We have studied the critical behavior at the critical temperature
$T=T_{c1}$ in a great detail and found the analogy between the critical
behavior at $T_{c1}$ and the van der Waals system. We also calculated the
critical exponents. The critical exponents $(0,\frac{1}{2},1,3)$ are the same
as that of mean field systems, which means that they are in the same
universality class. The corresponding scaling symmetry of free energy has also
been discussed. We finally plot the phase diagram of Kerr-AdS black hole.

The multi-critical phenomena we described in this paper could be a guide which
provides a restriction for a complete theory of underlying mechanism. We think
that the similar van der Waals-like structure, but the difference critical
exponents of RN-AdS and Kerr-AdS black holes deserves further study. The
natural extension, KN-AdS black hole, would be the next target to be understood.

Our results are interesting under the context of AdS/CFT duality. One can
study a strong correlation system dual to Kerr-AdS black hole. In
\cite{Sonner:2009fk}, Sonner found that superconducting-like condensation also
exists on the boundary of KN-AdS black hole. He found that there is a critical
value of rotation which could destroy the superconductivity in analogy to the
critical magnetic field. It is nature to guess that KN-AdS black hole has the
similar phase structure as Kerr-AdS and RN-AdS black holes, which will in some
way affect the thermal stability of the condensation. It would be interesting
to have further study along this direction.

In recent years, people managed to rewrite the field equations of gravity into
thermodynamics identities (for a review see \cite{Padmanabhan:2010xe}). In
\cite{Verlinde:2010hp}, Verlinde suggested that the gravity could be
understood as the entropy force induced by the inner freedom of the
holographic screen. These observations hope to regard gravity theory as a
result of more fundamental thermodynamic principles. It could be possible to
use the ideas to explain the thermal stability and phase structure of Kerr-AdS
black hole.

\begin{acknowledgments}
This work is supported in part by the National Science Council, 50 billions
project of Ministry of Education and National Center for Theoretical Science, Taiwan.
We thank Professor Tzay-Ming Hong and Professor Ling-Fong Li for fruitful discussions. 
\end{acknowledgments}

\bibliographystyle{plain}
\bibliography{Reference}

\end{document}